\pdfoutput=1
\documentclass[journal=jacs,manuscript=article,layout=twocolumn]{achemso}

\usepackage{graphicx}
\usepackage{dcolumn}
\usepackage{bm}
\usepackage{dsfont}
\usepackage{amsmath}
\usepackage{physics}
\usepackage{color}
\usepackage{soul}
\usepackage{url}
\usepackage[hidelinks]{hyperref}
\usepackage{scalerel}

\title{\Large  Mode-Selective and Anharmonicity-Controlled \\ Energy Transport in Cavity-Coupled Water}

\author{Sachith Wickramasinghe}
\affiliation{Department of Chemistry, Texas A\&M University, College Station, Texas 77843, USA}

\author{Michael Fowler}
\affiliation{Department of Chemistry, Texas A\&M University, College Station, Texas 77843, USA}

\author{Gerrit Groenhof}
\affiliation{Department of Chemistry, University of Jyväskylä, Finland}

\author{\\Arkajit Mandal}
\email{mandal@tamu.edu}
\affiliation{Department of Chemistry, Texas A\&M University, College Station, Texas 77843, USA}

\begin{document} 

\begin{abstract}
{\footnotesize
Recent experiments demonstrate the modification of chemical dynamics via cavity-enhanced vibrational energy transport. Here, we provide a microscopic account of both photonic and mode-selective energy transport using direct mesoscale on-the-fly simulations and provide the mechanistic principles of cavity-modified transport under vibrational strong coupling. We find that molecular anharmonicity plays a crucial role in dictating photonic transport, and driving-dependent photonic localization occurs when coupling cavity modes to a highly anharmonic mode of the molecular system. We demonstrate this in cavity-coupled water by tuning the photon frequency (at normal incidence) close to either the harmonic (or weakly anharmonic) bending mode or the anharmonic stretching modes of water. We confirm our understanding using a simple model system by reproducing the photonic transport and its localization. We also demonstrate that the diffusion of mode-selective temperature, quantified via the variance of the H-O-H bond angle or of the O-H bond length, is highly dependent on the cavity photon frequency. We show that the cavity photon frequency can be used as a tuning knob to achieve and control mode-selective energy transport. We also provide a simple analytical understanding of this phenomenon. Our results highlight the rich dynamical interplay of molecular and photonic degrees of freedom that persist in real atomistic systems.}
  
\end{abstract}
\maketitle
{\footnotesize
\section{\normalsize Introduction} 

Chemists have long regarded mode-specific control of reactions as a holy grail of chemistry.\cite{yin2025overcoming, morichika2019molecular, maas1998vibrational, nagarajan2021chemistry, simpkins2021mode} While synthetic strategies for achieving bond-selective reactions have been developed for decades and are continually being refined, attempts to realize the same goal via selective vibrational excitation with intense laser fields have failed to materialize as a generalized tool for chemical synthesis.~\cite{gruebele2004vibrational, morichika2019molecular} This is because any targeted vibrational energy rapidly leaks into other modes via intramolecular vibrational energy redistribution (IVR), which competes with the desired chemical pathway relating to the targeted vibrational excitation.~\cite{nesbitt1996vibrational, gruebele2004vibrational} This rapid redistribution arises from anharmonic coupling among the molecule’s normal modes, which enables fast vibrational energy flow through molecular bonds. In contrast, through-space vibrational energy transfer (between two molecules) is significantly slower and rarely plays a dominating role in determining chemical dynamics as it is mediated by substantially weaker intermolecular interactions.~\cite{xiang2020intermolecular}

Recent experiments, however, suggest a fundamentally different route: modification of ground-state chemical reactivity via the formation of vibropolaritons~\cite{nagarajan2021chemistry, thomas2019tilting, thomas2020ground, li2022molecular, mandal2023theoretical, thomas2016ground, ahn2023modification, yin2025overcoming, ruggenthaler2023understanding, campos2023swinging, lather2020improving, hasyim2026non} (light–matter hybrid quasi-particles) inside infrared (IR) optical cavities in the vibrational-strong-coupling (VSC) regime, where an ensemble of molecular vibrations is coupled to confined radiation fields. Specifically, recent experiments show that a driven VSC system, which is the focus of the present study, enables rapid through-space transport of vibrational excitations between spatially separated molecules over otherwise inaccessible distances and timescales, which has been suggested to modify chemical dynamics.~\cite{xiang2020intermolecular, yin2025overcoming} Interestingly, experiments also suggest the modification of chemical kinetics at thermal equilibrium~\cite{nagarajan2021chemistry, thomas2019tilting, thomas2020ground, vergauwe2019modification, LatherACIE2019, lather2021cavity}, even in the absence of external driving,~\cite{kenacohen2019polariton} albeit with some concerns about these claims due to subtleties in the interpretation of spectroscopic data.~\cite{vergauwe2026toward, Michon2024impact, imperatore2021reproducibility, wiesehan2021negligible} Nevertheless, cavity-enabled through-space transport has the potential to unlock the long-sought ability to perform mode-selective chemistry noninvasively and, when fully realized, could represent a paradigm shift in how we think about chemical synthesis. Importantly, the microscopic mechanistic principles of VSC-enabled transport and consequent cavity-modified chemical dynamics remain poorly understood. 

In this work, using direct on-the-fly mesoscale simulations, we provide a microscopic account of cavity-modified energy transport in water and establish fundamental mechanistic principles governing photonic transport and mode-selective, through-space vibrational energy transport inside an optical cavity. Specifically, we demonstrate how molecular anharmonicity controls photonic transport and how cavity photon frequency can be tuned to achieve mode-selective energy transport. We find that molecular anharmonicity modulates localization of photonic density, leading to a nonlinear dependence on the driving laser power. We show that when the cavity photon frequency at normal incidence is resonant with nearly harmonic molecular modes, such as the bending mode of water, which exhibits relatively weak anharmonicity, the photonic localization initially increases with pump power before plateauing at higher driving strengths. In contrast, when the cavity is resonant with the significantly more anharmonic water stretching modes, photonic localization exhibits a non-monotonic dependence on pump power. This mirrors our observation in many-body quantum dynamics of exciton-polaritons where the many-body excitonic interactions play a similar role as molecular anharmonicity.~\cite{GhoshPRB2025} We further demonstrate that tuning the cavity photon frequency can selectively enhance vibrational energy diffusion through either the bending or stretching modes. Together, these results establish mechanistic principles for understanding how molecular anharmonicity and cavity frequency dictate energy transport under VSC in realistic molecular systems.

While not the direct focus of this work, our work also sheds light on a fundamental conundrum of cavity modification of chemical reactivity~\cite{lindoy2024investigating, campos2023swinging, campos2020polaritonic}: how can collective cavity couplings, which connect cavity radiation to molecular degrees of freedom in a delocalized manner, with each individual molecule experiencing only a negligible fraction of the total coupling, nevertheless lead to modifications of chemical reactivity that operate locally? The cavity-enhanced energy transport investigated here operates in the collective-coupling regime. By enabling long-range, frequency-selective vibrational energy transport, the cavity provides a mechanism for translating a collective light–matter interaction into a cavity-controlled local modification of chemical reactivity. Such control directly addresses a central bottleneck in mode-selective chemistry: determining where vibrational energy flows on ultrafast timescales. The present work lays the foundation for future investigations into how cavity-modified energy diffusion can be harnessed to control chemical dynamics.}

\begin{figure*}[!t]
    \centering
    \includegraphics[width=1.0\linewidth]{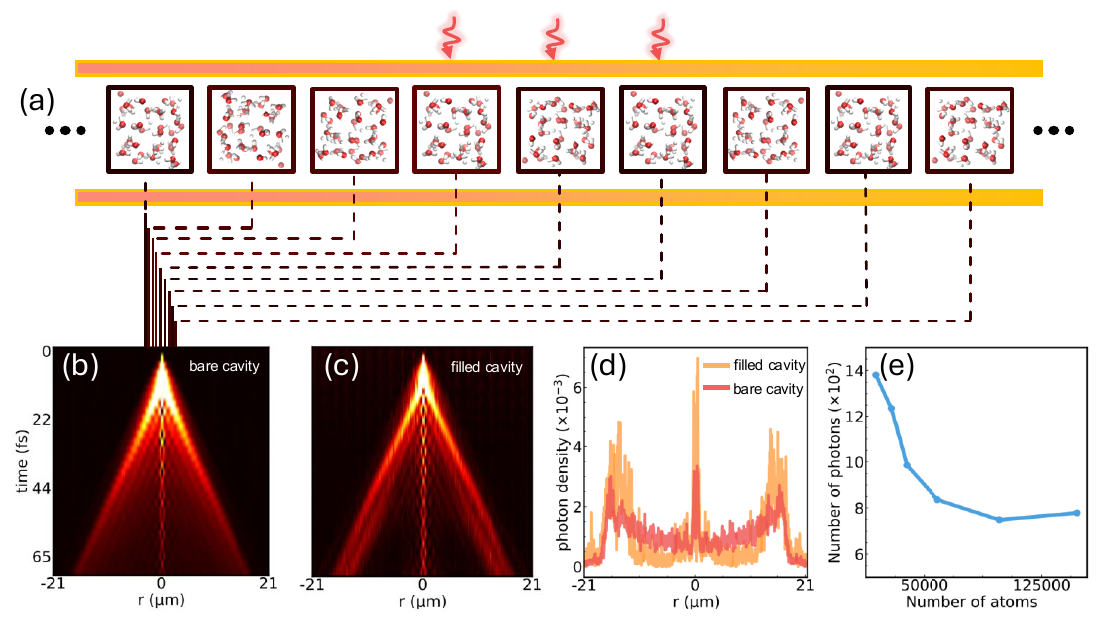} 
    \caption{\footnotesize \textbf{Mesoscale On-the-Fly Simulations of Cavity-Coupled Water and Their Convergence.} (a) Schematic representation of $N$ unit cells of water molecules aligned in an optical cavity with laser-driven central cells. (b) Propagation of photons over 72 fs within a bare cavity. (c) Propagation of polaritons over 72 fs within a cavity filled with coupled water molecules. (d) Photon density at 72 fs for the filled cavity (orange) and bare cavity (red). (e) Photon number convergence with system size. We set $g_0 = 2.2015 \times 10^{-5}$ a.u., $\omega_0 = 0.16$ eV, and $g_L = 0.6$ a.u. }
    \label{fig:1}
\end{figure*}

{\footnotesize 

\section{\normalsize Theory}

In this work, we perform mesoscale simulations using our recently developed cavOTF framework~\cite{wickramasinghe2026fly}, which is formulated in the dipole gauge beyond the long-wavelength approximation.~\cite{mandal2023microscopic, jasrasaria2025simulating, wickramasinghe2026fly, li2024vibrational, ji2026nonlinear} Specifically, we adapt a mixed representation (where parts of the Hamiltonian are represented in reciprocal space while the rest is represented in real space) of the light-matter Hamiltonian, written as~\cite{wickramasinghe2026fly} 
\begin{equation}\label{eq:hlm}
       \hat H_\mathrm{LM} = \hat H_\mathrm{M} + \sum_{k} \omega_k \hat a_k^\dagger \hat a_k  +   \eta \sum_{n} \hat q_n \hat{\mu}_n  +  \frac{\eta^2}{2 \omega_0^2} \sum_{n}  \hat{\mu}_n^2   .
\end{equation} 
Here, $\hat{H}_\mathrm{M}$ represents the molecular Hamiltonian which is written as a sum of $N$ non-interacting unit cells, following prior works,~\cite{wickramasinghe2026fly, ji2026nonlinear, li2024vibrational} such that 
\begin{align}
    \hat{H}_\mathrm{M} = \sum_{n} \hat{H}_{\mathrm{M}_n} = \sum_{n, j} \hat{T}_{n,j} + \hat{V}_{n}(\{r_{n,j}\}),
    \label{eq:hm}
\end{align}
where $\hat{T}_{n,j} $ is the kinetic energy operator of the $j$th atom in the $n$th unit cell, and $\hat{V}_{n}(\{r_{n,j}\})$ is the corresponding potential energy operator. In Eq.~\ref{eq:hlm}, the operators $\hat a_k^\dagger$ ($\hat a_k$) denote the photon creation (annihilation) operators corresponding to the $k$th cavity mode with frequency $\omega_k = c \sqrt{(\pi/L)^2+k^2}$, where $k = \frac{2\pi j}{L_s}$ ($j = 0, \pm 1, \pm 2, ...$) is the in-plane wavevector with $L_s$ as the length of the supercell, and $c$ is the speed of light. The parameter $\eta = \sqrt{\omega_0^3} g_0$ (with $g_0$ as the normalized coupling) denotes the collective light-matter coupling strength which has been set to place the system in the vibrational strong coupling regime, $\hat{q}_n = \frac{1}{\sqrt{2\omega_0}}(\hat{a}^{\dagger}_{n} + \hat{a}_{n})$ is the real space photonic position operator, $\omega_0$ is the cavity photon frequency at normal incidence, and $\hat \mu_n$ is the total dipole operator of the $n$th unit cell. We propagate the dynamics of the nuclear and photonic degrees of freedom classically, i.e. $\{\hat r_{n,j}, \hat q_n \} \rightarrow \{ r_{n,j}, q_n \}$. In our approach, we exploit the sparsity of the mixed real-reciprocal space description of the light-matter interaction, devising a low-communication hub-and-spoke server-client parallelization scheme to perform our mesoscale simulations.~\cite{wickramasinghe2026fly} The equations of motion for both the nuclear and photonic coordinates are derived from Hamilton’s equations of motion and integrated in time using a velocity-verlet-like scheme where we switch between real and reciprocal space in each nuclear timestep. Details of our propagation scheme can be found in Ref.~\citenum{wickramasinghe2026fly}. In addition to $\hat H_\mathrm{LM}$ in Eq.\ref{eq:hlm}, we include a time-dependent laser that drives the photonic modes, such that the total Hamiltonian, $\hat H_\mathrm{Total}$, is written as  
\begin{align} 
  \hat H_\mathrm{Total} =  \hat H_\mathrm{LM} +  \mathcal{E}(t)\sum_n g_{n} \hat{q}_n  .
    \label{eq:laser}
\end{align}
Here, $g_{n} = g_{L}\sum\limits_{m \in W} \delta_{nm}$, with $g_{L}$ as the laser driving strength and $W$ defines the spatial window over which the laser is focused, as illustrated in Fig.~\ref{fig:1}a. The temporal profile of the external field is given by $\mathcal{E}(t) = \Theta(t) \cdot [1 - \Theta(t - t_s)] \sin (\omega_l t)$, where $\Theta(t)$ is a Heaviside function, such that the laser is turned on between $t \in [0, t_s]$. We set $\omega_L = \omega_0$ and $t_s = 10$ fs; the short pulse duration produces a {\it broad} excitation spectrum. 

In our simulation setup, we consider an optical cavity with a total length of $L_s = 8 \times 10^{5}$ (a.u.). We place $N$ equally-spaced unit cells along the in-plane direction of the cavity. Each unit cell is a periodic cubic cell with a size of $10~{\text{\AA}}$ and contains 33 water molecules, matching the density of bulk water at room temperature.~\cite{wickramasinghe2023jpcb} We use DFTB+ to solve the electronic structure and compute the atomic forces, dipoles, and (numerical) dipole gradients on-the-fly. Initial positions and velocities of the water molecules were sampled from a canonical (NVT) molecular dynamics simulation. We used the `SupercellFolding' scheme\cite{hourahine2020dftb+} to perform electronic structure calculations with SCC-DFTB on a 3x3x3 folded supercell, where we used $\Gamma$-point sampling to approximate Brillouin zone integration. 

\begin{figure*}
\centering\includegraphics[width=1.0\linewidth]{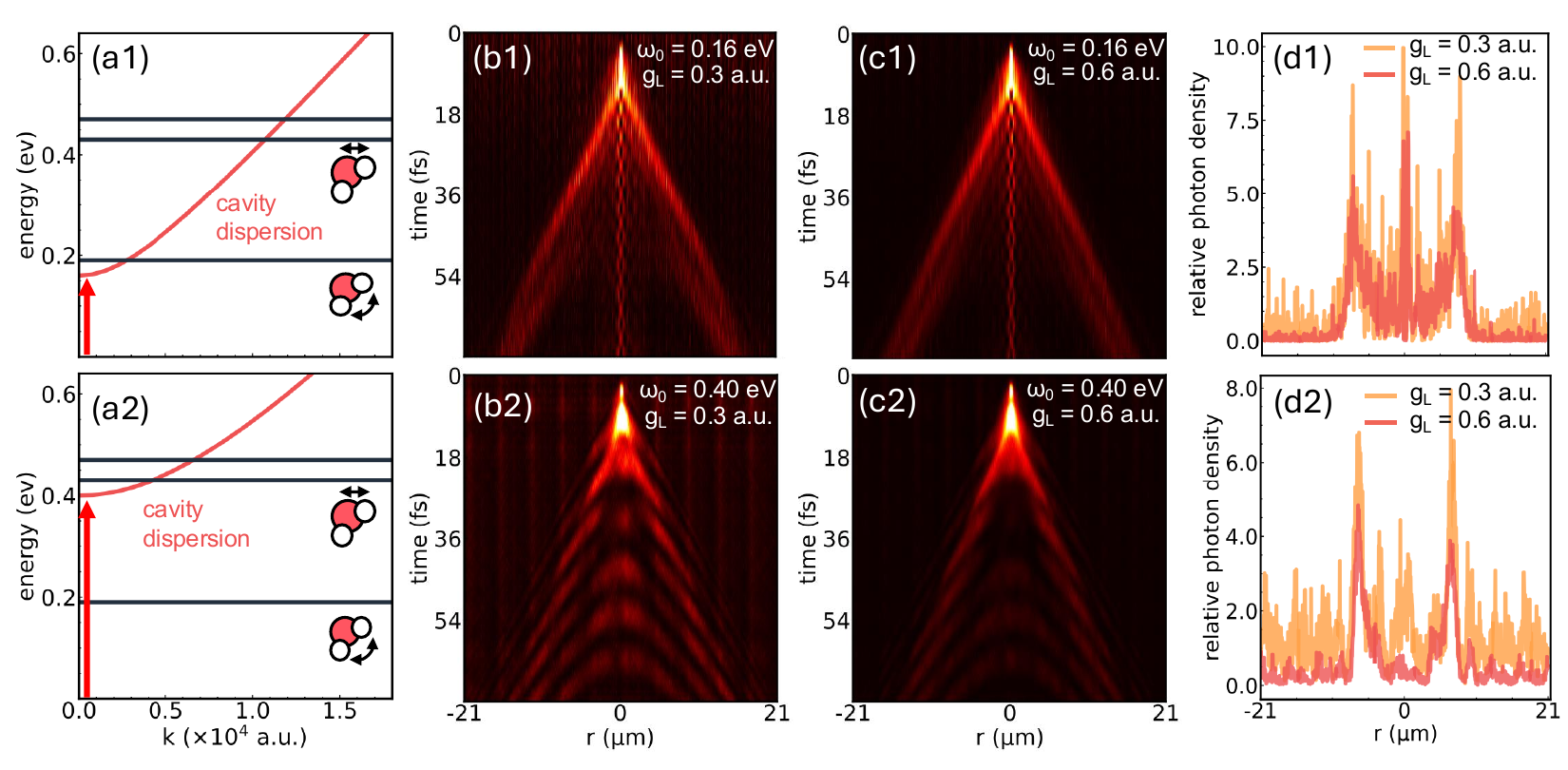} 
    \caption{\footnotesize \textbf{Anharmonicity-Controlled Transport.} (a) Dispersion of cavity (red) when excited at (a1) $\omega_0 = 0.16$ eV and (a2) $\omega_0 = 0.40$ eV and energies of water (black) bending ($\omega_\theta \sim 0.19$ eV) and stretching ($\omega_{v1} \sim 0.43$ eV and $\omega_{v2} \sim 0.47$ eV) modes. (b) Polaritonic propagation with driving strength of $g_L = 0.3$ a.u. when the cavity frequency is near resonance with (b1) the bending mode and (b2) the stretching modes. (c) Polaritonic propagation with driving strength of $g_L = 0.6$ a.u. when the cavity frequency is near resonance with (c1) the bending mode and (c2) the stretching modes. (d) Relative photon density at 72 fs at driving strengths $g_L = 0.3$ a.u. (orange) and $g_L = 0.6$ a.u. (red) when the cavity frequency is near resonance with (d1) the bending mode and (d2) the stretching modes. We set $g_0 = 2.2015 \times 10^{-5}$  a.u. in (a1)-(d1) and $g_0 = 1.8936 \times 10^{-5}$ a.u. in (a2)-(d2). }
    \label{fig:2}
\end{figure*}

\section{\normalsize Results and Discussion} 

\textbf{Mesoscale On-the-Fly Simulations of Cavity-Coupled Water and Their Convergence.} In Fig.~\ref{fig:1}, we present typical mesoscale simulations of vibrational polaritons formed in cavity-coupled water. Our setup is schematically illustrated in Fig.~\ref{fig:1}a. We set $W$, the excitation window, to 11 central unit cells. In Fig.~\ref{fig:1}b, we present the spatially-resolved photonic transport in an empty cavity and compare it with the photonic transport in a filled cavity (i.e., cavity-coupled water), shown in Fig.~\ref{fig:1}c, when setting $\omega_0 = 0.16$ eV. We compute the spatially-localized photon number at the $n$th unit cell classically as
\begin{equation}
    N_n(t) = \frac{1}{2}\left( \frac{p_n^2(t)}{\omega_0}+\omega_0\cdot\left( q_n(t)+\frac{\alpha \cdot\mu_n(t)}{\omega_0^2} \right)^2 \right),
    \label{eq:photonnumber}
\end{equation}
where $p_n$ is the photonic momentum (which relates to the vector potential), $q_n$ (which relates to the electric field) is the photonic position, $\omega_0$ is the cavity frequency, and $\alpha$ is a gauge parameter~\cite{stokes2019gauge} found by variationally minimizing the total photon number (which is done once for each set of parameters). We note that the choice of $\alpha$ does not alter the overall profile of the spatially-resolved photon number presented here.

Overall, here we observe both localized and ballistically propagating photonic density. This is because our broad laser excitation near the bottom of the photonic dispersion creates photonic density with both zero and non-zero group velocities. Coupling to water does change the time-dependent photonic density, as shown in Fig.~\ref{fig:1}d, in that a portion of the photonic density between the fast-propagating and fully localized parts is suppressed. This is due to the fact that the bending mode of water at $\omega_\theta = 0.19$ eV couples to the photonic density with intermediate group velocity.

Importantly, we find that the coupled molecular-photonic dynamics is sensitive to the number of unit cells $N$ considered in the simulation, even though the light-matter coupling is appropriately normalized. To ensure convergence, we have computed multiple observables of interest, one of which, the total number of photons at 72 fs at progressively higher number of unit cells (consequently higher number of total atoms, while keeping the number of atoms per unit cell constant), is presented in Fig.~\ref{fig:1}e. We find that $\sim 10^5$ atoms are needed to obtain reasonably converged results for the present setup, which we therefore use in our simulations.

\textbf{Anharmonicity-Controlled Transport.} Experimental results suggest that localization of polaritonic wavefunctions disrupts vibrational energy transport through the cavity.~\cite{yin2025overcoming} However, the present theoretical understanding of this phenomenon~\cite{liu2025unlocking} is limited to a single cavity mode model (employing the long-wavelength approximation), which cannot capture polaritonic transport, and uses a simple two-level molecular model that cannot capture rich molecular complexities. Using our on-the-fly simulation, we provide a microscopic account of how energy transport, quantified here via photonic transport, depends on molecular anharmonicity.

To elucidate the role of molecular anharmonicity in photonic transport, we consider two setups; we use a cavity with $\omega_0 = 0.16$ eV (Fig.~\ref{fig:2}a1-d1) near the bending mode at $\omega_\theta \sim 0.19$ eV, which is primarily harmonic in nature, and $\omega_0 = 0.40$ eV (Fig.~\ref{fig:2}a2-d2) near the stretching modes at $\omega_{v1} \sim 0.43$ eV and $\omega_{v2} \sim 0.47$ eV, which are significantly more anharmonic in nature. The uncoupled dispersions for these two setups are illustrated in Fig.~\ref{fig:2}a1-a2.

\begin{figure}[!t]
    \centering
    \includegraphics[width=1.0\linewidth]{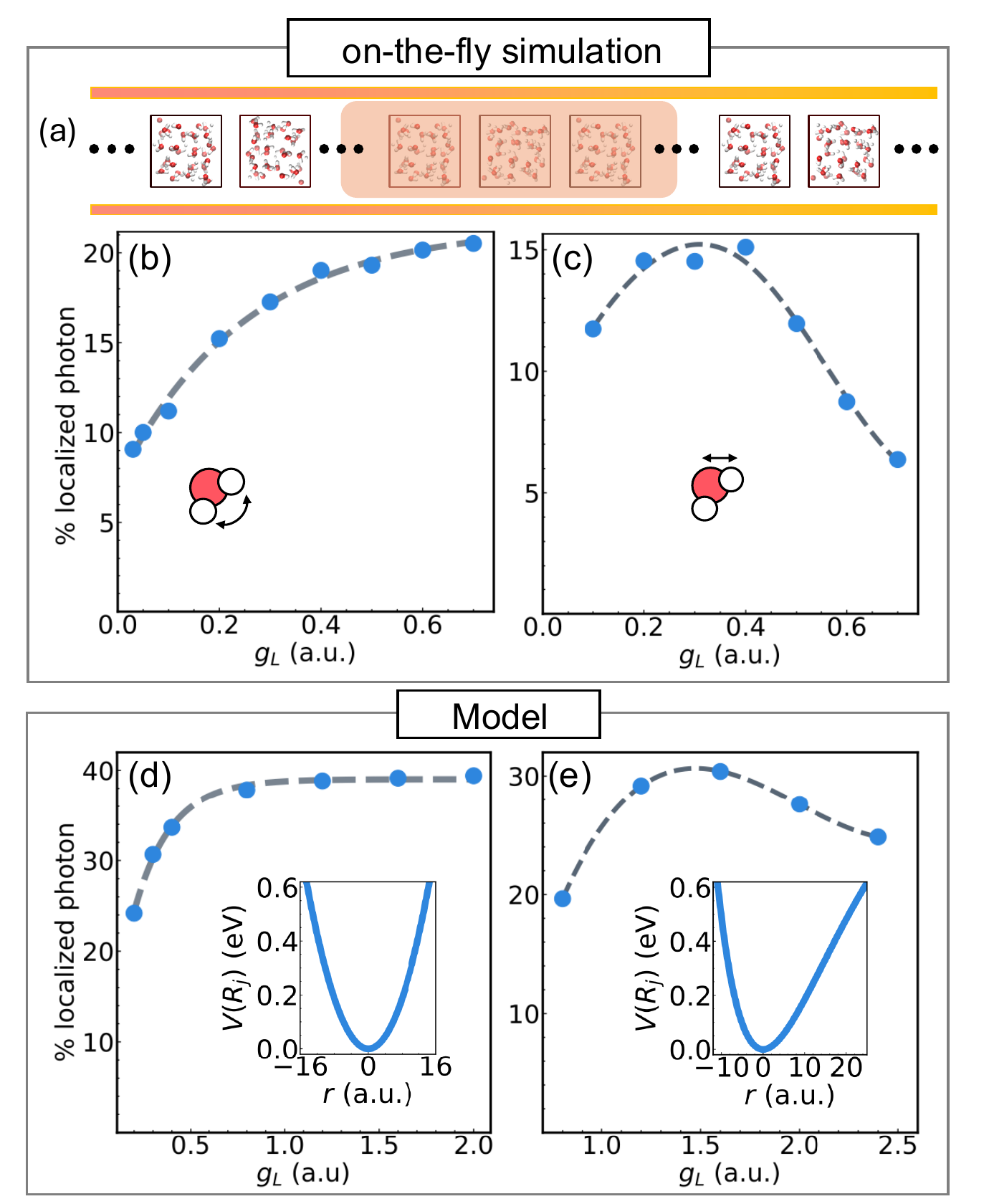} 
    \caption{\footnotesize  \textbf{Capturing Atomistic Results Using Reduced Models.} (a) Selected window for computing \% of localized photons is the center cell $\pm 50$, spanning $\sim$10\% of the total cavity length. (b) \% of localized photons in simulation when $\omega_0 = 0.16$ eV ($g_0 = 2.2015 \times 10^{-5}$ a.u.) at varying driving strengths $g_L \in (0.00, 0.70)$ a.u. (c) \% of localized photons in simulation when $\omega_0 = 0.40$ eV ($g_0 = 1.8936 \times 10^{-5}$ a.u.) at varying driving strengths $g_L \in (0.00, 0.70)$ a.u. (d) \% of localized photons when using a simple harmonic oscillator model system. (e) \% of localized photons when using a Morse potential model system.}
    \label{fig:3}
\end{figure}

In Fig.~\ref{fig:2}b1-c1, we present the spatially-resolved time-dependent photonic density at two different driving strengths $g_L$. Overall, we find that photonic transport is nearly identical when changing $g_L$ for $\omega_0 = 0.16$ eV, even though the total number of photons in the cavity setup scales as $g_L^2$, as expected. This can be further observed in Fig.~\ref{fig:2}d1 which presents the spatially-resolved relative photonic density, $N_n/g_L^2$, at 72 fs. The same result is also obtained when using simple harmonic models (as shown in Fig.~\ref{fig:3}); the photonic propagation is independent of the driving strength (beyond a thermal noise floor). 

Interestingly, the spatially-resolved time-dependent photonic density is drastically altered while increasing the photonic driving strength $g_L$ when setting $\omega_0 = 0.40$ eV, close to the anharmonic modes of water. Fig.~\ref{fig:2}b2-c2 demonstrates that the extent of the photonic localization is different at  $g_L = 0.3$ a.u. and $g_L = 0.6$ a.u. in that the localized portion disappears when using a relatively high photonic driving. This can be clearly observed in Fig.~\ref{fig:2}d2, which displays the relative photonic density at 72 fs. Note the localized photonic density near the center of the cavity ($r = 0$) at $g_L = 0.3$ a.u. (orange line) which is missing at $g_L = 0.6$ a.u. (red line). These results also mirror the result obtained in many-body quantum dynamics of exciton-polaritons in the presence of exciton-exciton onsite interaction terms which can be viewed as anharmonicity.~\cite{GhoshPRB2025} Note that molecular anharmonicity, modeled as a Morse potential, corresponds to an attractive many-body interaction term in the Bose-Hubbard picture. Due to this many-body attraction (which increases with increasing photon number), propagating photonic density is expected to pull away localized densities and suppress localization. Classically, a larger extent of driving is expected to dynamically tune the effective  frequency of the anharmonic mode, which allows for the photonic population trapped near $k\sim 0$ to scatter to  higher $k\ne 0$ modes with finite velocity, making this photonic density mobile. This mechanism is consistent with our observations.

\begin{figure*}[!t]
    \centering
    \includegraphics[width=1.0\linewidth]{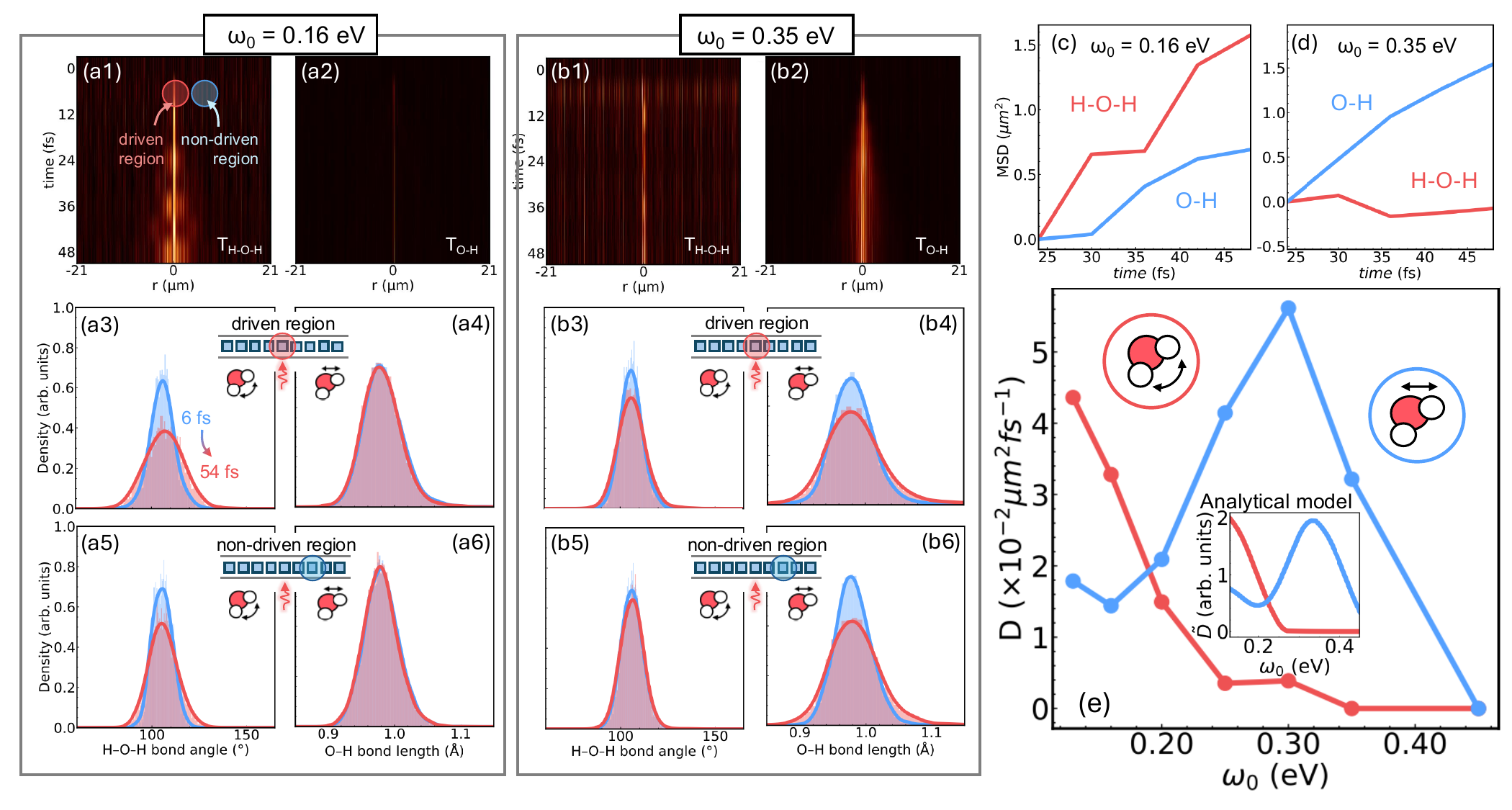} 
    \caption{\footnotesize \textbf{Mode-Selective Energy Transport.} (a) Cavity frequency $\omega_0 = 0.16$ eV, near bending mode. (a1) Effective temperature $T_\mathrm{H-O-H}$ diffusion over 54 fs. (a2) Effective temperature $T_\mathrm{O-H}$ diffusion over 54 fs. (a3, a4) Distributions of (a3) bond angles and (a4) bond lengths in the driven region at $t = 6$ fs and $t = 54$ fs. (a5, a6) Distributions of (a5) bond angles and (a6) bond lengths in the non-driven region at $t = 6$ fs and $t = 54$ fs. (b) Cavity frequency $\omega_0 = 0.35$ eV, near stretching modes. (b1) Effective temperature $T_\mathrm{H-O-H}$ diffusion over 54 fs. (b2) Effective temperature $T_\mathrm{O-H}$ diffusion over 54 fs. (b3, b4) Distributions of (b3) bond angles and (b4) bond lengths in the driven region at $t = 6$ fs and $t = 54$ fs. (b5, b6) Distributions of (b5) bond angles and (b6) bond lengths in the non-driven region at $t = 6$ fs and $t = 54$ fs. (c, d) MSDs of bending mode (red) and stretching mode (blue) energy diffusion at (c) $\omega_0 = 0.16$ eV and (d) $\omega_0 = 0.35$ eV. (e) Cavity frequency-dependent mode-resolved energy diffusion and comparison to analytical model. We set $g_0 = 1.8936 \times 10^{-5}$ a.u.}
    \label{fig:4}
\end{figure*}

To unambiguously demonstrate that the suppression of photonic localization originates from molecular anharmonicity, we compute photonic localization, $\mathcal{N}_{loc}$, as a function of photonic driving and compare it with a reduced model system. We define photonic localization as the ratio of photons found in the middle of the cavity versus the entire cavity setup, expressed as

\begin{equation}
    \mathcal{N}_{loc} = \frac{\int_{t-\Delta t}^{t+\Delta t}d\tau\int_{-r_c}^{+r_c} dr N(r, \tau)}{\int_{t-\Delta t}^{t+\Delta t}d\tau\int_{-\infty}^{+\infty} dr N(r, \tau)},
\end{equation}

where $N(r, \tau) \approx \sum_{n} N_n(\tau) \cdot \delta(r - a\cdot n)$, and $r_c = 2.11 ~\mu$m sets the bounds of the localized region (corresponding to $\pm 50$ unit cells) which is $\sim$10\% of the total cavity length. In the following, we set $t = 21$ fs and use a finite $\Delta t = 6$ fs to reduce noise. We note that this localization is unrelated to the recently observed nonlinear freezing of photonic density.~\cite{ji2026nonlinear}  

Fig.~\ref{fig:3}b presents the \% photonic localization ($\%\mathcal{N}_{loc}$) as a function of driving strength $g_L$ when setting $\omega_0 = 0.16$ eV, close to the harmonic bending mode of water. At thermal equilibrium, under no driving, $\%\mathcal{N}_{loc}$ is expected to be $\sim$10\%, consistent with the localized window occupying $\sim$10\% of the total cavity length. This explains the $\%\mathcal{N}_{loc}$ as $g_{L} \rightarrow 0$ in Fig.~\ref{fig:3}b. As $g_{L}$ is increased, the localized photonic density overcomes the thermal noise, and we consequently observe an increase in the \% photonic localization in Fig.~\ref{fig:3}b. Importantly, once $g_{L}$ is sufficiently large (i.e., $g_{L}> 0.4$ a.u.), the amount of photonic localization saturates.

The same is observed in Fig.~\ref{fig:3}d when using a simple model system where the molecular degrees of freedom are described as a collection of harmonic oscillators. That is, we replace $\hat{H}_\mathrm{M_n} \rightarrow \hat{H}_\mathrm{M_n}^{model} =  \sum_j \frac{p_{j,n}^2}{2} + V(R_j)$ with $V(R_j) = \frac{1}{2}\Omega^2 R_{j}^2$, a harmonic potential (illustrated in the inset of Fig.~\ref{fig:3}d), while we keep the rest of the terms identical in $\hat{H}_\mathrm{Total}$. Further details can be found in the supporting information.
As observed in Fig.~\ref{fig:3}b, the photonic localization saturates at some $g_{L}$. 

Fig.~\ref{fig:3}c presents the \% photonic localization ($\%\mathcal{N}_{loc}$) as a function of driving strength $g_L$ when setting $\omega_0 = 0.40$ eV, close to the anharmonic stretching modes of water. Importantly, we observe that the photonic localization nonlinearly depends on the photonic driving; it increases at smaller values of $g_L$ (similar to Fig.~\ref{fig:3}b) but then starts to decrease with increasing $g_L$, indicating a suppression of localization as was discussed in Fig.~\ref{fig:2}.

Fig.~\ref{fig:3}e confirms that this nonlinear effect originates from molecular anharmonicity. Here, we consider a model molecular system where the molecular degrees of freedom are described as anharmonic oscillators, i.e., we set $V(R_j) = \frac{\Omega}{4 \chi_e}\big(1-e^{-\sqrt{2 \Omega \chi_e}\cdot R_{j}} \big)^2$, a Morse potential (illustrated in the inset of Fig.~\ref{fig:3}e), where $\chi_e = 0.08$ a.u. is the anharmonicity parameter (where $\chi_e \rightarrow 0$ means $V(R_j) \rightarrow \frac{1}{2}\Omega^2 R_{j}^2$ ). The $g_L$-dependent photonic localization presented in Fig.~\ref{fig:3}e clearly reproduces the nonlinear, non-monotonic dependence of $\mathcal{N}_{loc}$ on $g_L$, which was absent in Fig.~\ref{fig:3}d. Overall, our results presented in Fig.~\ref{fig:2}-\ref{fig:3} clarify how molecular anharmonicity controls (multi-) photonic dynamics. We emphasize that these dynamical insights can only be captured when employing a mesoscale beyond long-wavelength description~\cite{li2024vibrational, wickramasinghe2026fly, ji2026nonlinear, haines2026mechanistic, jasrasaria2025simulating, rahmanian2026exciton, tichauer2021multi, blackham2025microscopic, Krupp2025, sokolovskii2024one, tichauer2023tuning} and that a single cavity mode limit adapted in recent theoretical works~\cite{li2021cavity, li2022qm, li2021collective, lindoy2023quantum, mondal2026macroscopic, li2021cavity_jcp, du2023vibropolaritonic,sun2024theoretical, sun2023modification, hasyim2026toward} is not adequate.

\textbf{Mode-Selective Energy Transport.} In Fig.~\ref{fig:4}, we now focus on how molecular dynamics is also impacted due to light-matter coupling. Specifically, our results show that coupling to cavity photons enables mode-selective vibrational energy transport which can be tuned by altering the cavity mirror spacing (i.e., by changing $\omega_0$). 

Fig.~\ref{fig:4}a-b presents the mode-selective vibrational energy transport at two different $\omega_0$. To quantify the vibrational energy transport, we compute the H-O-H bond angle and O-H bond length distributions. The corresponding probability densities (in arb. units) are presented in Fig.~\ref{fig:4}a3-a6 and Fig.~\ref{fig:4}b3-b6. From these distributions, we compute a spatially-resolved {\it effective} temperature (i.e., for each unit cell), defined as $T_\mathrm{H-O-H} = \langle \theta^2 \rangle - \langle \theta \rangle^2$ and $T_\mathrm{O-H} = \langle r_\mathrm{O-H}^2 \rangle - \langle r_\mathrm{O-H} \rangle^2$, where $\theta$ is the H-O-H bond angle and $r_\mathrm{O-H}$ is the O-H bond length. This particular choice of  {\it effective} temperature follows from the fact that for perfectly harmonic potentials, such as $\frac{1}{2}\Omega^2R^2$, the classical Boltzmann distribution leads to the relationship $T \propto \langle R^2 \rangle - \langle R \rangle^2$. Note that while the O–H stretching potential is anharmonic, the bond length variance still provides a convenient measure of spatial redistribution of the stretching excitation.

Fig.~\ref{fig:4}a1-a2 presents the spatially-resolved time-dependent effective temperature when setting $\omega_0 = 0.16$ eV. The time-dependent $T_\mathrm{H-O-H} $ heatmap presented in Fig.~\ref{fig:4}a1 shows that the unit cells progressively get `heated', indicating energy transport, after a short-time excitation (which persists for the initial 10 fs) in the central region of the cavity with $r \in [-0.234 ~\mu m,0.234 ~\mu m]$. In contrast, we observe significantly slower transport in Fig.~\ref{fig:4}a2 which presents the time-dependent $T_\mathrm{O-H}$ heatmap. 

A microscopic view is presented in Fig.~\ref{fig:4}a3-a6. Fig.~\ref{fig:4}a3 presents the density of the bending angle $\rho_{\theta}$ within the {\it driven} unit cells with $r \in [-0.234 ~\mu m,0.234 ~\mu m]$. Clearly, $\rho_{\theta}$ is broadened over time, indicating the deposition of energy into the bending mode of water. Note that the bending mode is not directly coupled to the external laser, but it acquires energy indirectly from the cavity photon modes which are directly driven. Within the same time window, we observe little broadening of $\rho_{O-H}$ (density of the O-H bond length), presented in Fig.~\ref{fig:4}a4. Obviously, at much longer times, $\rho_{O-H}$ will also broaden due to intramolecular IVR. Meanwhile, in Fig.~\ref{fig:4}a5, we present the density of the bending angle $\rho_{\theta}$  within the {\it non-driven} unit cells (neighboring the driven unit cells), with $r \in [0.255 ~\mu m,0.468 ~\mu m]$. We observe that $\rho_{\theta}$ is broadened over time, indicating a cavity-enabled energy transfer. The same is not true in Fig.~\ref{fig:4}a6 which shows marginal changes in $\rho_{O-H}$. 

When setting $\omega_0 = 0.35$ eV, which is closer to the stretching modes, the results presented in Fig.~\ref{fig:4}b1-b6 show the opposite trend compared to Fig.~\ref{fig:4}a1-a6. Here, the propagation of $T_\mathrm{O-H}$ (Fig.~\ref{fig:4}b2) is more substantial than the propagation of $T_\mathrm{H-O-H}$ (Fig.~\ref{fig:4}b1). This can also be observed by analyzing the microscopic picture presented in Fig.~\ref{fig:4}b3-b6. Specifically, we see that both the driven and non-driven regions show a substantial broadening of $\rho_{O-H}$ (compare Fig.~\ref{fig:4}b4 and b6). In contrast, while the driven region displays a substantial broadening of $\rho_{\theta}$ (Fig.~\ref{fig:4}b3), which is mediated by IVR, the broadening of $\rho_{\theta}$ in the non-driven region (Fig.~\ref{fig:4}b5) is much less so, suggesting a much slower propagation of $T_\mathrm{H-O-H}$.

In Fig.~\ref{fig:4}c-d, we present the mean square displacement (MSD) of $T_\mathrm{H-O-H}$ and $T_\mathrm{O-H}$. The MSDs are calculated as

\begin{equation}
    \mathrm{MSD}(t) = \frac{\int_{-\infty}^{\infty}dr~r^2 T_j(r,t + t_s)}{\int_{-\infty}^{\infty}dr  T_j(r,t + t_s)} - \frac{\int_{-\infty}^{\infty}dr~r^2 T_j(r, t_s)}{\int_{-\infty}^{\infty}dr  T_j(r, t_s)},
\end{equation}

where $T_j \in [T_\mathrm{H-O-H}, T_\mathrm{O-H}]$, and $t_s = 24$ fs is a reference time. The MSDs for $T_\mathrm{H-O-H}$ and $T_\mathrm{O-H}$ at $\omega_0 = 0.16$ eV are presented in Fig.~\ref{fig:4}c. It is clear that in the present scenario, the MSD for $T_\mathrm{H-O-H}$ is larger than for $T_\mathrm{O-H}$. In contrast, the MSDs presented in Fig.~\ref{fig:4}d show the opposite behavior, $T_\mathrm{O-H}$ propagates much faster compared to $T_\mathrm{H-O-H}$ at $\omega_0 = 0.35$ eV, corroborating our findings in Fig.~\ref{fig:4}a-b. 

Finally, to illustrate the frequency-dependent and mode-selective nature of vibrational energy transport, quantified by $T_\mathrm{O-H}$ and $T_\mathrm{H-O-H}$, we compute an (effective) diffusion constant from our MSDs as $D = \frac{\mathrm{MSD}(t_f)}{2t_f}$, where we set $t_f = 24$ fs (corresponding to $t = t_f + t_s = 48$ fs). Note that the MSDs presented here are not perfectly linear; nevertheless, this $D$ allows for comparing the relative propagation of the effective temperatures.  Our mesoscale simulations show that the diffusion of $T_\mathrm{O-H}$ and $T_\mathrm{H-O-H}$ can be tuned by changing $\omega_0$. For low $\omega_0$ ($<0.2$ eV), we find the diffusion of $T_\mathrm{O-H}$ to be much smaller than that of $T_\mathrm{H-O-H}$, which switches at $\omega_0\sim0.2$ eV so that the diffusion constant for $T_\mathrm{O-H}$ becomes much larger than the diffusion constant for $T_\mathrm{H-O-H}$. At much larger $\omega_0$ ($\sim 0.45$ eV), both diffusion constants become negligible. 

This cavity frequency dependence of mode-selective `energy' diffusion can be rationalized using a {\it crude} analytical picture. Quantum mechanically, using Markovian Redfield dynamics, the diffusion constant is written as~\cite{yarkony1977variational, munn1973direct, cheng2008unified}

\begin{equation}
    D_\mathrm{QM} \approx \sum_{k} \frac{v_k^2}{\Gamma_k}\rho_{kk}, 
\end{equation}

where $\rho_{kk}$ is the population of the $k$th state (where $k$ is the wavevector), $v_k = d\omega_k/dk$ is the group velocity, and $\Gamma_k$ is a scattering rate. When assuming a $k$-independent scattering rate, the diffusion constant is then proportional to the weighted average of the group velocity (i.e., $D_\mathrm{QM} \propto \sum_k v_k^2 \rho_{kk}$). Noting that the number of photons at $k$ is expected to be proportional to $|\mathcal{E}(\omega_k)|^2$ (where $\mathcal{E}(\omega)$ is the sine-transformed $\mathcal{E}(t)$ in Eq.~\ref{eq:laser}) and that the transition rate from photon to molecular vibrations is proportional to the oscillator strength $I$, we arrive at the proportionality $\rho_{kk} \propto I\cdot |\mathcal{E}(\omega_k)|^2$. Following this, we arrive at a crude analytical expression 

\begin{equation}
    \tilde{D}(\omega) = \frac{I}{\Delta \omega} \int_{\omega- \Delta\omega/2}^{\omega+ \Delta\omega/2} d\omega v_k^2(\omega) \cdot |\mathcal{E}(\omega)|^2,
\end{equation}

where $\Delta\omega = 50$ meV accounts for a finite energy window which is populated, and we treat $I$ as a free parameter (set to 1 for the H-O-H bending mode and 2.2 for the O-H stretching mode) which simply scales the overall line shape. Here, $\tilde{D}(\omega)$ should be considered proportional to the observed diffusion constant. We note that $v_k(\omega)$, $\mathcal{E}(\omega)$, and consequently $\tilde{D}(\omega)$ depend on $\omega_0$. We present $\tilde{D}(\omega_{\theta})$ and $\tilde{D}(\omega_{v_1})$ as functions of $\omega_0$ in the inset of Fig.~\ref{fig:4}e. Despite the crude nature of this analytical expression, we find that our analytical model qualitatively reproduces the simulated $\omega_0$-dependent diffusion constants.  This analysis reveals that energy transport associated with a particular molecular mode can be achieved on sub-picosecond timescales by bringing the photonic dispersion close to resonance with that mode. This enables rapid, cavity-mediated through-space energy transport over propagation lengths reaching the micrometer scale.

\section{Conclusion}

In this work, we have studied the microscopic picture of the through-space energy transport in a cavity-coupled water system via anharmonic and mode-selective control. Using our mesoscale on-the-fly atomistic simulations, we provide mechanistic principles for cavity-enabled energy transport under VSC which extend to the experimental scale. Our results highlight the importance of going beyond a simple model molecular description of the matter subsystem and the need for adapting a beyond long-wavelength description of light-matter interactions to understand cavity-enabled dynamical phenomena. 

We show that photonic transport is sensitive to the coupled molecular mode and its anharmonicity. We find that highly anharmonic molecular modes suppress localization of photonic density, especially at higher photonic driving. In contrast, the localization of photonic density becomes nearly independent of photonic driving when coupling to a harmonic molecular mode. We demonstrate this in cavity-coupled water, where the bending mode is harmonic (or weakly anharmonic), but the stretching modes are relatively anharmonic. We use reduced models to verify this mechanistic principle. Finally, we show that photon frequency at normal incidence can be used to tune the relative diffusion of mode-selective propagation of {\it effective} temperature (consequently energy transport), quantified here by the spatially-resolved time-dependent variances of the bond angle and bond length distributions of water. We show that this mode-selectivity can be understood as a consequence of the interplay between photonic group velocity and frequency-dependent laser intensity.  

Overall, our work sheds light on the complex light-matter dynamics under VSC and provides fundamental mechanistic insights. Our future work will focus on how mode-selective transport may lead to mode-selective chemical dynamics.

\section{Acknowledgments}
This work was supported by the Texas A\&M startup funds. This work used TAMU ACES and LAUNCH clusters at the Texas A\&M University through allocation PHY230021 and CHE250162 from the Advanced Cyberinfrastructure Coordination Ecosystem: Services \& Support (ACCESS) program, which is supported by National Science Foundation grants \#2138259, \#2138286, \#2138307, \#2137603, and \#2138296. We acknowledge the EuroHPC Joint Undertaking for awarding this project access to the EuroHPC supercomputer LUMI, hosted by CSC (Finland) and the LUMI consortium through a EuroHPC Regular Access call.



\bibliography{ref.bib}

 }
\end{document}